\documentclass[useAMS,usenatbib]{mn2e}
\usepackage{mn2e-breakabs}
\usepackage{graphicx}
\usepackage{times}
\usepackage{hyperref}
\def\fun#1#2{\lower3.6pt\vbox{\baselineskip0pt\lineskip.9pt
        \ialign{$\mathsurround=0pt#1\hfill##\hfil$\crcr#2\crcr\sim\crcr}}}

\newcommand{\be}{\begin{equation}}
\newcommand{\ee}{\end{equation}}
\newcommand{\ba}{\begin{eqnarray}}
\newcommand{\ea}{\end{eqnarray}}
\newcommand{\simgt}{\,\hbox{\lower0.6ex\hbox{$\sim$}\llap{\raise0.6ex\hbox{$>$}}}\,}
\newcommand{\simlt}{\,\hbox{\lower0.6ex\hbox{$\sim$}\llap{\raise0.6ex\hbox{$<$}}}\,}

\begin{document}

\title[Photometric Redshift Estimator for SNe~Ia from LSST]
{Analytic Photometric Redshift Estimator for Type Ia Supernovae\\ From the Large Synoptic Survey Telescope}

\author[Wang, Gjergo, \& Kuhlmann]{
  \parbox{\textwidth}{
    Yun Wang$^{1,2}$\thanks{E-mail: wang@ipac.caltech.edu}, E. Gjergo$^3$, and S. Kuhlmann$^3$}
  \vspace*{4pt} \\
$^1$Infrared Processing and Analysis Center, California Institute of Technology,
770 South Wilson Avenue, Pasadena, CA 91125\\
  $^2$Homer L. Dodge Department of Physics \& Astronomy, Univ. of Oklahoma,
                 440 W Brooks St., Norman, OK 73019, U.S.A.\\
  $^3$Argonne National Laboratory, 9700 South Cass Avenue, Lemont, IL 60439, USA
                 }

\date{\today}

\maketitle

\begin{abstract}
Accurate and precise photometric redshifts (photo-z's) of Type Ia supernovae (SNe Ia)
can enable the use of SNe Ia, measured only with photometry, to probe cosmology.
This dramatically increases the science return of supernova surveys planned for 
the Large Synoptic Survey Telescope (LSST).
In this paper we describe a significantly improved version of the simple analytic  
photo-z estimator proposed by Wang (2007) and further developed by Wang, Narayan, and Wood-Vasey (2007).
We apply it to 55,422 simulated SNe Ia generated using the SNANA package with the LSST filters.
We find that the estimated errors on the photo-z's, $\sigma_{z_{\rm phot}}/(1+z_{\rm phot})$, can be used
as filters to produce a set of photo-z's that have high precision, accuracy, and purity.
Using SN Ia colors as well as SN Ia peak magnitude in the $i$ band, we obtain a set of
photo-z's with 2 percent accuracy
(with $\sigma(z_{\rm phot}-z_{\rm spec})/(1+z_{\rm spec}) = 0.02$), a bias in $z_{\rm phot}$
(the mean of $z_{\rm phot}-z_{\rm spec}$) of $-9\times 10^{-5}$, and an outlier
fraction (with $\left|(z_{\rm phot}-z_{\rm spec})/(1+z_{\rm spec})\right|>0.1$) of
0.23 percent, with the requirement that $\sigma_{z_{\rm phot}}/(1+z_{\rm phot})<0.01$.
Using the SN Ia colors only, we obtain a set of photo-z's with similar quality by
requiring that $\sigma_{z_{\rm phot}}/(1+z_{\rm phot})<0.007$; this leads to a set
of photo-z's with 2 percent accuracy, a bias in $z_{\rm phot}$ of $5.9\times 10^{-4}$,
and an outlier fraction of 0.32 percent.

\end{abstract}


\section{Introduction}
\label{sec:intro}

The use of Type Ia supernovae (SNe Ia) as cosmological standard candles
is a corner stone of modern cosmology, and led to the observational discovery of cosmic acceleration \citep{Riess,Perl}.
The true nature of cosmic acceleration, dubbed ``dark energy'' for convenience, remains shrouded in mystery.
We do not even know if it is an unknown form of energy (hence appropriately named ``dark energy''), or a consequence of the modification of 
Einstein's theory of general relativity (``modified gravity'').\footnote{For recent reviews, see 
\cite{Ratra08,Frieman08,Caldwell09,Uzan10,Wang10,Li11,Weinberg12}.}
Probing the true nature of cosmic acceleration requires the measurement
of both cosmic expansion history and the growth history of large scale structure \citep{Knox06,Guzzo08,Wang08}. 
The use of SNe Ia remains one of the most important methods for measuring cosmic expansion history.

We can expect a dramatic increase in the number of SNe Ia that can be used
to measure cosmic expansion history in the coming years and decades.
Thousands of SNe Ia are expected from the Dark Energy Survey (DES) \citep{bern12}, and 
hundreds of thousands of SNe Ia are expected from Large Synoptic Survey Telescope
(LSST) \citep{lsst}.\footnote{\url{http://www.lsst.org/lsst/science}}

The majority of the SNe Ia from LSST will only have photometry, and 
will not have spectroscopic redshifts, due to the expensive resources required to obtain 
follow-up spectroscopy. Fortunately, the redshifts of the SNe Ia can be estimated
using multi-band photometry (these approximate redshifts are called photometric redshifts,
or photo-z's). However, the observed photometric SNe need to be classified first using 
photometry (see ``Supernova Photometric Classification Challenge'' by \cite{Kess10b} for a detailed discussion). 
Even applying an optimized version of the top performing method from the Supernova Photometric Classification Challenge,
the photometric-classification algorithm of \cite{Sako11} result in over 25\% of the resultant photometric SN Ia 
sample remaining non-Ia SNe \citep{Campbell13}.\footnote{The photometric-classification method of \cite{Campbell13} is based on
the SN classification technique of \cite{Sako11}, but made additional cuts aided by host-galaxy redshifts with $0.05 < z < 0.55$. This enabled them to achieve 
the SN Ia classification efficiency of 70.8\%, with only 3.9\% contamination from core-collapse SNe.}
We will defer the difficult task of photometric classificationo future work, and focus on the relatively easier 
task of estimating redshifts of known SNe Ia using photometry only.

SNe~Ia with sufficiently accurate and precise photo-z's can be used for probing cosmology. 
This would dramatically increase the science return 
of supernova surveys planned for LSST.
Accurate photo-z's can also enhance the
ability of observers to accurately target high redshift SNe~Ia for spectroscopy,
allowing flexibility and optimization in survey design for Stage III and 
Stage IV dark energy projects \citep{DETF}.

In this paper, we build on the previous work by \cite{Wang07} and
\cite{Wangetal2007}, and present a simple analytic photo-z estimator. We apply 
this photo-z estimator to 55,422 simulated SNe Ia generated using the SNANA package \citep{snana} with the LSST filters.
We present our method in Sec.\ref{sec:method},
describe the simulation of LSST SNe Ia in Sec.\ref{sec:sims}, show our results
in Sec.\ref{sec:results}, and discuss our findings in Sec.\ref{sec:sum}.

\section{The Method}
\label{sec:method}

The analytic photo-$z$ estimator for SNe~Ia proposed by \citet{Wang07}
is empirical, model independent (no templates used), 
and uses observables that reflect the properties
of SNe~Ia as calibrated standard candles.
It was developed using the SN~Ia data released by the Supernova
Legacy Survey \citep{Astier06}.

This method requires a training set of SNe Ia with spectroscopic redshifts. 
The SNe Ia in the training set are the same as the data, but with known redshifts.
The training set is randomly chosen from the data to span the ranges of the colors
and magnitudes of the entire data set. This mimics what should be done in practice when
real data become available: choose a training set from the data to carry out spectroscopic
redshift measurements.

\subsection{The Prototype Photo-z Estimator}

The prototype photo-z estimator we use here was developed
in Wang (2007) and Wang, Narayan, and Wood-Vasey (2007).
This estimator uses the fluxes in $griz$ (or $riz$) at the epoch 
of $i$ maximum flux to make an effective K-correction
to the $i$ flux. The first estimate of redshift is given by
\be
z_{\rm phot}^{0}=c_1 + c_2 g_f +c_3 r_f + c_4 i_f + c_5 z_f +c_6 i_f^2
+ c_7 i_f^3
\label{eq:z0}
\ee
where $g_f=2.5\log(f_g)$, $r_f=2.5\log(f_r)$, $i_f=2.5\log(f_i)$,
and $z_f=2.5\log(f_z)$, and $f_g$, $f_r$, $f_i$, $f_z$ are fluxes
in counts, normalized to some fiducial zeropoint, in $griz$ at the epoch of $i$ maximum flux.

Next, the photo-z estimator calibrates each SN~Ia in its estimated rest-frame using
\be
\Delta i_{15}= 2.5 \log(f_i^{15d}/f_i),
\label{eq:del_i15}
\ee
where $f_i^{15d}$ is the $i$ band flux at 15 days after
the $i$ flux maximum in the estimated rest-frame, corresponding
to the epoch of $\Delta t^{15d}=15 (1+z_{\rm phot}^{0})$ days after the
epoch of $i$ flux maximum.	    

The final estimate for the photometric redshift is given by
\be
z_{\rm phot} = \sum_{i=1}^8 c_i \,p_i,
\label{eq:z_a}
\ee
where the data vector ${\mbox {\bf p}}=\{1, g_f, r_f, i_f, z_f, i_f^2,
i_f^3, \Delta i_{15}\}$.
The coefficients $c_i$ (i=1,2,...,8) are found by using
a training set of SNe~Ia with $griz$ (or $riz$, for which $c_2=0$) 
lightcurves and
measured spectroscopic redshifts. 
The jackknife technique \citep{Lupton93} is used to estimate the 
mean and the covariance matrix of $c_i$.

\subsection{Photo-z Estimator Applied to LSST Filters}
\label{subsec:LSSTall}

For application to SNe Ia from LSST with photometry only, we
have simplified and modified the analytic photo-z estimator for easier application
and better performance as follows:\\
\noindent
(1) Instead of using the fluxes in all filters except i at the epoch 
of $i$ maximum flux to make an effective K-correction
to the $i$ flux, we just use the peak magnitudes in all filters
for each SN to construct an indicator of the spectral shape of the SN.\\
\noindent
(2) We add 2nd order terms in the colors for improved accuracy and
precision of the estimator.\\
\noindent
(3) We replace $\Delta i_{15}$ with $\Delta i_{12}=2.5 \log(f_i^{12d}/f_i)$, to increase the sample
size, where where $f_i^{12d}$ is the $i$ band flux at 12 days after
the $i$ flux maximum in the estimated rest-frame, corresponding
to the epoch of $\Delta t^{12d}=12 (1+z_{\rm phot}^{0})$ days after the
epoch of $i$ flux maximum.	    

Note that our use of $\Delta i_{15}$ was first introduced by \cite{Wang07} to reduce the scatter in the estimated 
SN Ia photometric redshifts; it is analogous to the use of $\Delta m_{15}$ in SN Ia calibration, 
but the use of $i$ band magnitude allows us to include a larger set of SNe Ia in our analysis \citep{Wang07}.

The key to obtaining a set of reliable photo-z's for SNe is to exclude
the SNe that are inferred to have unreliable photo-z estimates based
on the photometry data alone.
We make a first cut in SNe to be used as follows:\\
\noindent
(1) Exclude the few SNe that have no $i$-band photometry. This is
necessary since we use the $i$ magnitude at maximum light in
estimating the photo-z's.\\
\noindent
(2) Exclude SNe that have measured flux uncertainties at maximum light in any passband  
that exceeds 0.2 magnitude. This is useful in reducing the number of
outliers with $\left|(z_{\rm phot}-z_{\rm spec})/(1+z_{\rm spec})\right|>0.1$.\\
\noindent
(3) Exclude SNe that do not have $i$-band lightcurve data that
extends to 12 days after maximum light in the estimated SN restframe.
This enables the refinement of the photo-z estimated using the
fluxes at maximum light alone.

We then apply the photo-z estimator as follows:\\
\noindent
(1) Divide the SNe according to the passbands in which the
SN magnitudes at maximum light are available.\\
\noindent
(2) For each set of SNe with photometry in the same passbands,
construct the initial photo-z estimate given by
\ba
z_{\rm phot}^{0}&=&c_1 + \sum_{j=1}^{n-1} c_{j+1} (m_{p,j}-m_{p,j+1})\nonumber\\
&&+\sum_{j=1}^{n-1} c_{n+j} (m_{p,j}-m_{p,j+1})^2 \nonumber\\
 & & +c_{2n} i_p + c_{2n+1}{i_p}^2 + c_{2n+2} {i_p}^3
\label{eq:z0new}
\ea
where $m_{p,j}$ is the magnitude at maximum light in the
j-th passband, $i_p$ is the magnitude at maximum light in the
$i$ band, $n$ is the number of filters or passbands with available photometry,
and $\{c_j\}$, $j=1,2, ..., 2n+2$, are the
coefficients to be estimated from using the training set
of SNe with spectroscopic redshifts.\\
\noindent
(3) Next, calibrate each SN~Ia in its estimated rest-frame using
\be
\Delta i_{12}= i_{12d}-i_p
\label{eq:del_i12}
\ee
where $i_{12d}$ is the $i$ band magnitude at 12 days after
the $i$ maximum light in the estimated rest-frame, corresponding
to the epoch of $\Delta t^{12d}=12 (1+z_{\rm phot}^{0})$ days after the
epoch of $i$ maximuml light.	    

The final estimate for the photometric redshift is given by
\be
z_{\rm phot} = \sum_{j=1}^{2n+3} c_j \,p_j,
\label{eq:za_new}
\ee
where the data vector ${\mbox {\bf p}}=\{1, 
(m_{p,1}-m_{p,2}), (m_{p,2}-m_{p,3}), ..., (m_{p,n-1}-m_{p,n}), 
(m_{p,1}-m_{p,2})^2, (m_{p,2}-m_{p,3})^2, ..., (m_{p,n-1}-m_{p,n})^2, 
i_p, i_p^2, i_f^3, \Delta i_{12}\}$.
The coefficients $c_j$ ($j=1,2,...,2n+3$) are found by using
a training set of SNe~Ia with lightcurves in $n$ passbands and
measured spectroscopic redshifts. 
We use the same modified jackknife technique as in Wang, Narayan, and Wood-Vasey (2007)
to estimate the mean and the covariance matrix of $c_j$ (see Sec.~\ref{sec:results}).

The improvements that we have made to the photo-z estimator in this work reduces the 
estimated errors on the photo-z's, $\sigma_{z_{\rm phot}}/(1+z_{\rm phot})$, from 2.5\% to 2\%,
compared to \cite{Wangetal2007}.
More importantly, it reduces the bias in $z_{\rm phot}$ (the mean of $z_{\rm phot}-z_{\rm spec}$),
from $-1.5 \times 10^{-3}$ to $-9\times 10^{-5}$ (a reduction factor of 16.67).

\subsection{Photo-z Estimator Using Colors Only}
\label{subsec:colors_only}

Since our ultimate goal is to use SNe Ia with photometry to probe cosmology,
it is desirable to have a photo-z estimator that does {\it not} use the
$i$ band flux information, to reduce the likelihood of possible double counting of distance information.
Note that SN Ia colors are also used in fitting for cosmology, one may argue that including the 
$i$ band flux information is no more double-counting than including SN Ia colors.
However, the $i$ band flux of a SN Ia is closely correlated with its distance from us, since
we are using the brightness of SNe Ia as distance indicators. 
SN color is also used in the distance estimation, but only for the correction to the distance, so it is a much smaller effect
compared to the SN brightness. Finally, the estimation of photometric redshifts are made possible
by the use of colors, thus they are essential ingredients in any photometric redshift estimator.
Therefore, having an estimator for SN Ia photometric redshifts that uses colors only (the minimal ingredients) enables a more robust
approach to probing cosmology using photometric SNe Ia.

We arrive at a photo-z estimator using colors only simply by omitting the last 4 terms in
Eq.(\ref{eq:za_new}).
Thus the simplified photo-z estimator that uses colors only is given by
\ba
z_{\rm phot}^{c}&=&c_1 + \sum_{j=1}^{n-1} c_{j+1} (m_{p,j}-m_{p,j+1})\nonumber\\
&&+\sum_{j=1}^{n-1} c_{n+j} (m_{p,j}-m_{p,j+1})^2 
\label{eq:zc}
\ea
where $m_{p,j}$ is the magnitude at maximum light in the
j-th passband, $n$ is the number of filters or passbands with available photometry,
and $\{c_j\}$, $j=1,2, ..., 2n-1$, are the
coefficients to be estimated from using the training set
of SNe with spectroscopic redshifts.

\section{Simulation of data}
\label{sec:sims}

The number and location of the LSST deep drilling fields are still to be determined; for a general discussion,
see the LSST Science Book.\footnote{http://www.lsst.org/lsst/scibook}
Assuming a plausible scenario of 10 fields covering 90 square degrees,
we obtained a set of 55,422 simulated SNe with photometric data including magnitude uncertainties,
generated using the SNANA package \citep{snana} with the LSST filters.
We have assumed 10 years of observations on the LSST deep-drilling fields.
We have used measured rates of SNIa \citep{Dilday08} as input to the simulations, and imposed the 
requirement that three LSST filters have a SNR greater than 5 (which gives two color measurements
essential for photo-z estimates).

We have applied 0.005 mag of smearing to the bandpass zeropoints; this simulates the 
variations due to the atmosphere. We do not include LSST Photon Simulator
effects or systematic bandpass zeropoint shifts at present.\footnote{The LSST Photon 
Simulator (see http://lsstdesc.org/WorkingGroups/PhoSim)
includes a very detailed simulation of the LSST hardware,
and takes a star or galaxy location and brightness and propagates the light through
the optics and camera. } 
These effects are expected to be smaller than the variations due to the atmosphere.

To exclude incomplete or very noisy data, we make a few basic cuts for data quality.
Only 55,414 of the 55,422 simulated SNe have peak i mags, and only 49,993 have $i$ band lightcurves.
Requiring that the $i$ lightcurve data include info of the $i$ peak
and extend to 12 days in the estimated SN restframe, and that all the
peak magnitudes have errors less than 0.2 magnitudes, we arrive at a
sample of 29,702 simulated SNe. The data quality cuts that we have made here are less
strict that those made by current surveys; nevertheless we find them to be sufficient 
because of the higher quality photometry expected from the LSST.
We will use this sample of 29,702 SNe Ia that have passed our data quality cut in the remainder of this paper.

We now describe our Type Ia SN light curve
simulations in greater technical detail. 
We employ the SNANA package \citep{snana} to simulate 
the light curves using the SALT2 Type Ia SN model. \footnote{The SALT2 model we are using is the most recently updated model available, with the latest data.}
The SNANA package was first used by the LSST
collaboration to forecast SN observations~\citep{lsst}, 
and has been used extensively by the SDSS collaboration~\citep{kes09}, 
to forecast observations for the Dark Energy Survey~\citep{bern12}, and
for many other systematic studies for supernova cosmology. 
The specific SALT2 model used is the extended version of the 
Guy 2010 model~\citep{guy07,guy10}.  
The SALT2 model rest-frame flux $F(p,\lambda)$ as a function of phase $p$ and wavelength $\lambda$, 
is given in equation~\ref{eqn:flux}.  The two spectral time series, $M_0$ 
and $M_1$, and the color law $CL$ are part of the specific SALT2 input model.
The parameters $x_0, x_1, c$, are randomly drawn from distributions that have
been fit previously to Type Ia SN data including the effects of dust extinction.  
Given these parameters the 
rest-frame flux as a function of phase is available for use.  The redshift is 
likewise randomly drawn from a previously measured distribution,  and the 
observer-frame flux is integrated with the LSST filter transmissions. 
\begin{equation}
 F(p,\lambda) = x_{0}[M_{0}(p,\lambda)+x_{1}M_{1}(p,\lambda)]e^{-cCL(\lambda).}
\label{eqn:flux}
\end{equation}

Our sample of simulated LSST SNe Ia is representative of data expected from the LSST,
based on our current knowledge of SNe Ia.
The SNANA simulations of SNe Ia have been heavily tested on SDSS photometric data up to $z<0.7$ (see,
e.g., \cite{Hlozek12,Campbell13}) ,  and
on SNLS data up to $z<1.0$.   This is similar to the redshift range we are using.

\section{Results}
\label{sec:results}

\subsection{Photo-z's using SN Ia colors and $i$ band peak flux}

We use a slightly modified version of the jackknife technique 
\citep{Lupton93} to estimate the 
mean and the covariance matrix of $c_j$. 
From the training set containing $N$ SNe~Ia, 
we extract $N$ subsamples each containing $N-1$ SNe~Ia
by omitting one SN~Ia. The coefficients $c_j^{(s)}$ ($j=1,2,...,2n+3$)
for the $s$-th subsample are found by a maximum likelihood
analysis matching the predictions of
Eq.~(\ref{eq:za_new}) with the spectroscopic redshifts.

The mean of the coefficients $c_j$ ($j=1,2,...,2n+3$) are given by
\be
\langle c_j \rangle = \frac{1}{N} \sum_{s=1}^{N} c_j^{(s)}.
\label{eq:mean}
\ee
Note that this is related to the usual ``bias-corrected jackknife
estimate'' for $c_i$, $c_i^J$, as follows:
\be
c_j^J \equiv c_j^N +(N-1) \left(c_j^N-\langle c_j \rangle\right),
\ee
where $c_j^N$ are estimated from the entire training set (with $N$ SNe~Ia).
\cite{Wangetal2007} found that for small training sets (with $N<20$) that include
SNe~Ia at $z \sim 0$, $c_j^J$ give
biased estimates of $c_j$ by giving too much weight to the
SN~Ia with the smallest redshift. 
For training sets not including nearby SNe~Ia, $\langle c_j \rangle$ and $c_j^J$ 
are approximately equal.
We have chosen to use $\langle c_j \rangle$ from Eq.~(\ref{eq:mean})
as the mean estimates for $c_j$ to avoid biased $z_{\rm phot}$ 
for SNe~Ia at $z$ close to zero \citep{Wangetal2007}.

The covariance matrix of $c_j$ ($j=1,2,...,2n+3$) are given by
\be
{\rm Cov}(c_i,c_j)= \frac{N-1}{N} \sum_{s=1}^{N} 
\left(c_i^{(s)} - \langle c_i \rangle \right)\, 
\left(c_j^{(s)} - \langle c_j \rangle \right)
\ee
Using $z_{\rm phot}= \sum c_j p_j$, we find that
$\Delta z_{\rm phot}= \sum p_j\Delta c_j$,
since the uncertainty in $z_{\rm phot}$ is dominated by the
uncertainty in $c_j$.
Therefore estimated error in $z_{\rm phot}$ is
\be
{\rm d}z_{\rm phot}= \left\{ \sum_{i=1}^{2n+3} \sum_{j=1}^{2n+3} 
 [p_i] \, {\rm Cov}(c_i,c_j) [p_j] \right\}^{1/2},
\ee
where ${p_i}$ and ${p_j}$ ($i,j=1,2,...2n+3$) are
the $i$-th and $j$-th components of the data vector
where the data vector ${\mbox {\bf p}}=\{1, 
(m_{p,1}-m_{p,2}), (m_{p,2}-m_{p,3}), ..., (m_{p,n-1}-m_{p,n}), 
(m_{p,1}-m_{p,2})^2, (m_{p,2}-m_{p,3})^2, ..., (m_{p,n-1}-m_{p,n})^2, 
i_p, i_p^2, i_f^3, \Delta i_{12}\}$.

The simulated SNe are divided into groups according to their
photometric properties, see Table \ref{tab:sets}.
Each training set for a group of SNe consists of the first 100 SNe
in that group, assuming that the simulated data are randomly ordered.
For groups that contain very small numbers of SNe, we use the first 20
SNe in each group as the training set, or increase the number of SNe
used in the training set until no negative $z_{\rm phot}$ results. 
This led us to a training set containing a total of 1040 SNe used for
fitting the analytic photo-z estimator, and a ``test set'' of 28,662 SNe 
that are {\it not} used in the fitting, and are used as a validation set
for the analytic photo-z estimator.

Table \ref{tab:sets} shows that we need ~1000 SN Ia spectra for the 
training sets for the 29,702 simulated SNe Ia that satisfy our minimal data quality cuts
(see Sec.3). When the real LSST photometric SN Ia data are available, we can choose 
the training set in each color group described in Table \ref{tab:sets} in a random fashion,
while ensuring that the SNe Ia chosen for the training set spans the required color and
magnitude range. Then spectroscopic redshifts can be obtained for the SNe Ia
in the training sets.

Even 1000 SN Ia spectra will require substantial observational resources.
The sizes of the training sets can be reduced to meet the constraints of the available resources, which may
degrade the accuracy and precision of the photometric redshifts.
However, the performance of the photometric redshifts can be recovered when the training sets are expanded
later via SN Ia host galaxy redshift measurements.
Two strategies can be used to reduce the number of SN Ia spectra needed for the training sets:
(1) Remove the SNe Ia belonging to the color groups containing very small fractions of the overall
sample (G16, G22, G61, and perhaps G21 in Table \ref{tab:sets}). Note that these minority color groups require
a much larger fraction of spectroscopic redshifts, and have less well estimated photo-z's.
Dropping these will save resources (100 less spectroscopic redshifts) with minimal impact on the cosmological constraints.
(2) Reduce the sizes of the training sets of the other color groups as needed. 

In applying the photo-$z$ estimator to simulated data, we find
that for each group of SNe with a given set of available
photometric passbands, applying a single fitting formulae to its training set
may lead to substructure in the estimated $z_{\rm phot}$. Therefore, we have
divided each group into subgroups as shown in Table \ref{tab:sets}, and derived the
photo-z estimator for each subgroup using its own training set.

Table \ref{tab:sets} tabulates the resultant precision and accuracy  of our photo-z estimator for
each color group. We measure precision with $\sigma(z_{\rm phot}-z_{\rm spec})/(1+z_{\rm spec}$,
and the accuracy by the mean of $z_{\rm phot}-z_{\rm spec}$, as well as the fraction of outliers (also
listed are the number of outliers that are flagged as having large uncertainties in $z_{\rm phot}$).

\begin{figure}
\includegraphics[width=\columnwidth,clip]{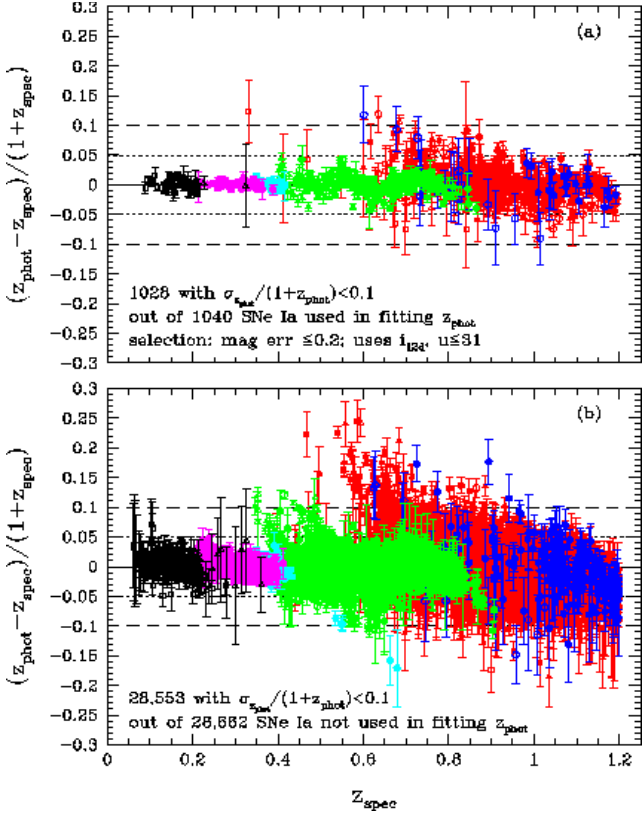}
\caption{The performance of the analytic photo-z estimator applied to simulated SNe from the LSST,
as indicated by $(z_{\rm phot}-z_{\rm spec})/(1+z_{\rm spec})$ versus $z_{\rm spec}$, with cut 
$\sigma_{z_{\rm phot}}/(1+z_{\rm phot})<0.1$ for cleaner presentation (the numbers of outliers are 
given in Table \ref{tab:sets})..
Different colors indicate SNe with different photometric passbands, 
as described in Table \ref{tab:sets}.
For SNe represented by the same color, different point types represent
subdivisions based on brightness to improve precision.}
\label{fig:zacut0d1}
\end{figure}

\begin{figure}
\includegraphics[width=1.2\columnwidth,clip]{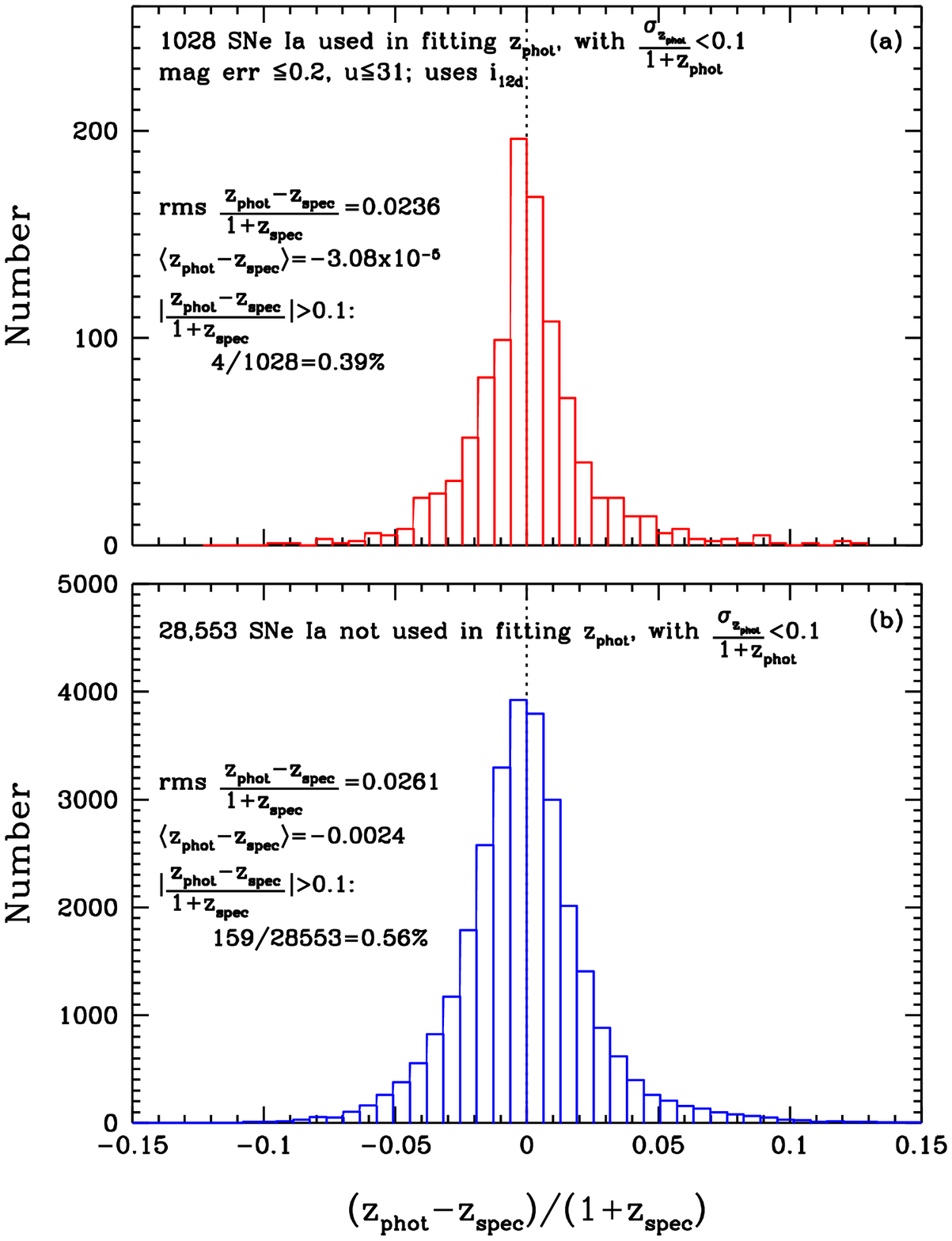}
\caption{The distribution of $(z_{\rm phot}-z_{\rm spec})/(1+z_{\rm spec})$ from Fig.\ref{fig:zacut0d1}.}
\label{fig:zacut0d1_bin}
\end{figure}

Fig.\ref{fig:zacut0d1} shows $(z_{\rm phot}-z_{\rm spec})/(1+z_{\rm spec})$ versus $z_{\rm spec}$
for the analytic photo-z estimator applied to simulated SNe from the LSST, 
with cut $\sigma_{z_{\rm phot}}/(1+z_{\rm phot})<0.1$.
The cut is made for better presentation; only 12 SNe out of 1040 of the training set,
and 109 out of 28662 of the test set have $\sigma_{z_{\rm phot}}/(1+z_{\rm phot})\geq 0.1$.
Different colors indicate SNe with different photometric passbands, 
as described in Table \ref{tab:sets}.
For SNe represented by the same color, different point types represent
subdivisions based on brightness to improve precision.
The histogram of the number of SNe versus $(z_{\rm phot}-z_{\rm spec})/(1+z_{\rm spec})$ 
corresponding to Fig.\ref{fig:zacut0d1} is shown in Fig.\ref{fig:zacut0d1_bin}.

\begin{figure}
\includegraphics[width=1.4\columnwidth,clip]{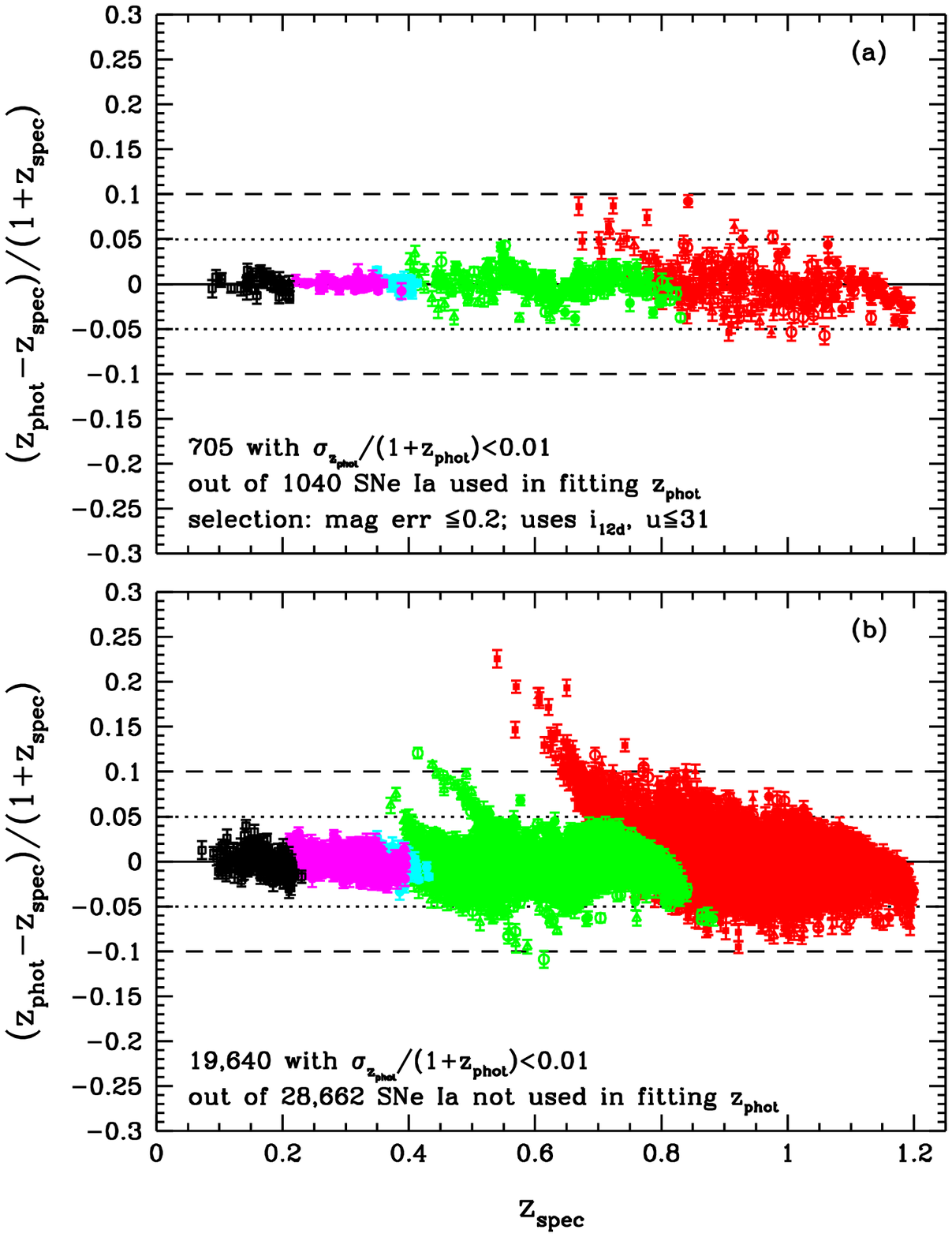}
\caption{The same as Fig.\ref{fig:zacut0d1}, but with cut 
$\sigma_{z_{\rm phot}}/(1+z_{\rm phot})<0.01$ for improved accuracy, precision, and outlier reduction.}
\label{fig:zacut0d01}
\end{figure}

\begin{figure}
\includegraphics[width=1.2\columnwidth,clip]{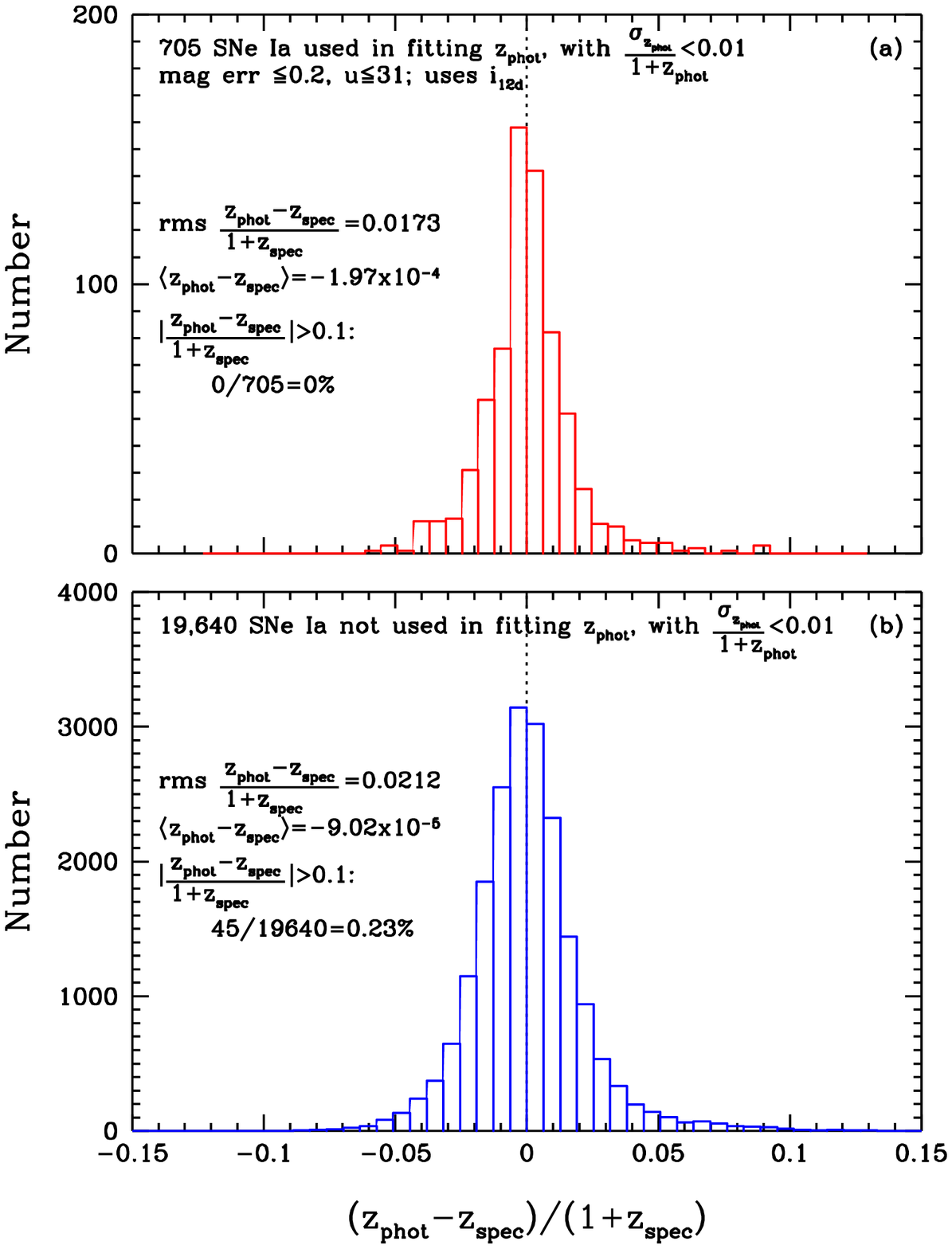}
\caption{The distribution of $(z_{\rm phot}-z_{\rm spec})/(1+z_{\rm spec})$ from Fig.\ref{fig:zacut0d01}.}
\label{fig:zacut0d01_bin}
\end{figure}

Fig.\ref{fig:zacut0d01} and Fig.\ref{fig:zacut0d01_bin} are the same as Fig.\ref{fig:zacut0d1} and
Fig.\ref{fig:zacut0d1_bin}, but for imposing the cut $\sigma_{z_{\rm phot}}/(1+z_{\rm phot})<0.01$.
There are 45 outliers (with $\left|(z_{\rm phot}-z_{\rm spec})/(1+z_{\rm spec})\right|>0.1$) 
in the bottom panel in Fig.\ref{fig:zacut0d01} (the majority are from the rizY color groups
G11 through G15),  representing 0.23\% of all 19,640 SNe Ia from the test sample that passed this cut.

Table \ref{tab:cuts} shows that we can arrive at photo-z's with higher accuracy, precision, and purity
by using the estimated errors on $z_{\rm phot}$ to exclude SNe with large $\sigma_{z_{\rm phot}}/(1+z_{\rm phot})$.
If we exclude SNe with $\sigma_{z_{\rm phot}}/(1+z_{\rm phot})\geq 0.01$, we arrive at a precision of
$\sigma(z_{\rm phot}-z_{\rm spec})/(1+z_{\rm spec}) = 0.02$, a bias in $z_{\rm phot}$
(the mean of $z_{\rm phot}-z_{\rm spec}$) of $-9\times 10^{-5}$, and an outlier
fraction (with $\left|(z_{\rm phot}-z_{\rm spec})/(1+z_{\rm spec})\right|>0.1$) of
0.23 percent (see Fig.\ref{fig:zacut0d01}). Note that for the threshold of $\sigma_{z_{\rm phot}}/(1+z_{\rm phot})< 0.01$,
there are no outliers in the training set. We can require that there are no outliers
in the training set in order to arrive at this threshold for $\sigma_{z_{\rm phot}}/(1+z_{\rm phot})$.

\subsection{Photo-z's using SN Ia colors only}
\label{subsec:colors_only_results}

As we discussed in Sec.\ref{subsec:colors_only}, having an alternative photometric redshift estimator that
uses SN Ia colors only is helps increase the robustness of cosmological constraints obtained using photometric
SNe Ia.
Table \ref{tab:sets} tabulates the resultant precision and accuracy  of our photo-z estimator using SN Ia colors only, for
each color group. The colors only results are tabulated immediately below the results for colors plus $i$ band peak flux
(see the previous subsection), and indicated by ``Y" under the column for ``colors only''.
Using SN colors only leads to slightly worse precision, but similar accuracy for the resultant photo-z's.

\begin{figure}
\includegraphics[width=\columnwidth,clip]{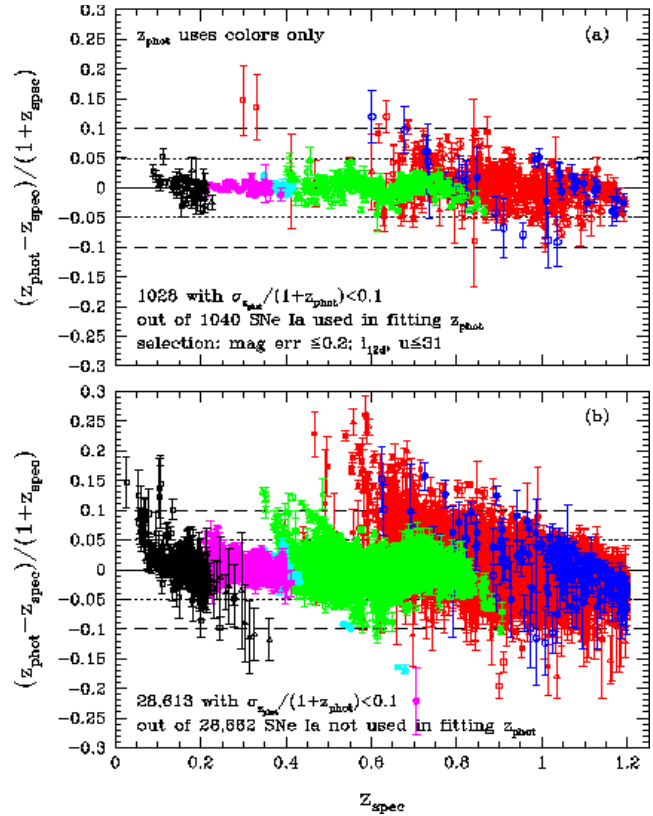}
\caption{The same as Fig.\ref{fig:zacut0d1}, but for $z_{\rm phot}$ obtained using SN Ia colors only.}
\label{fig:zaccut0d1}
\end{figure}

\begin{figure}
\includegraphics[width=1.4\columnwidth,clip]{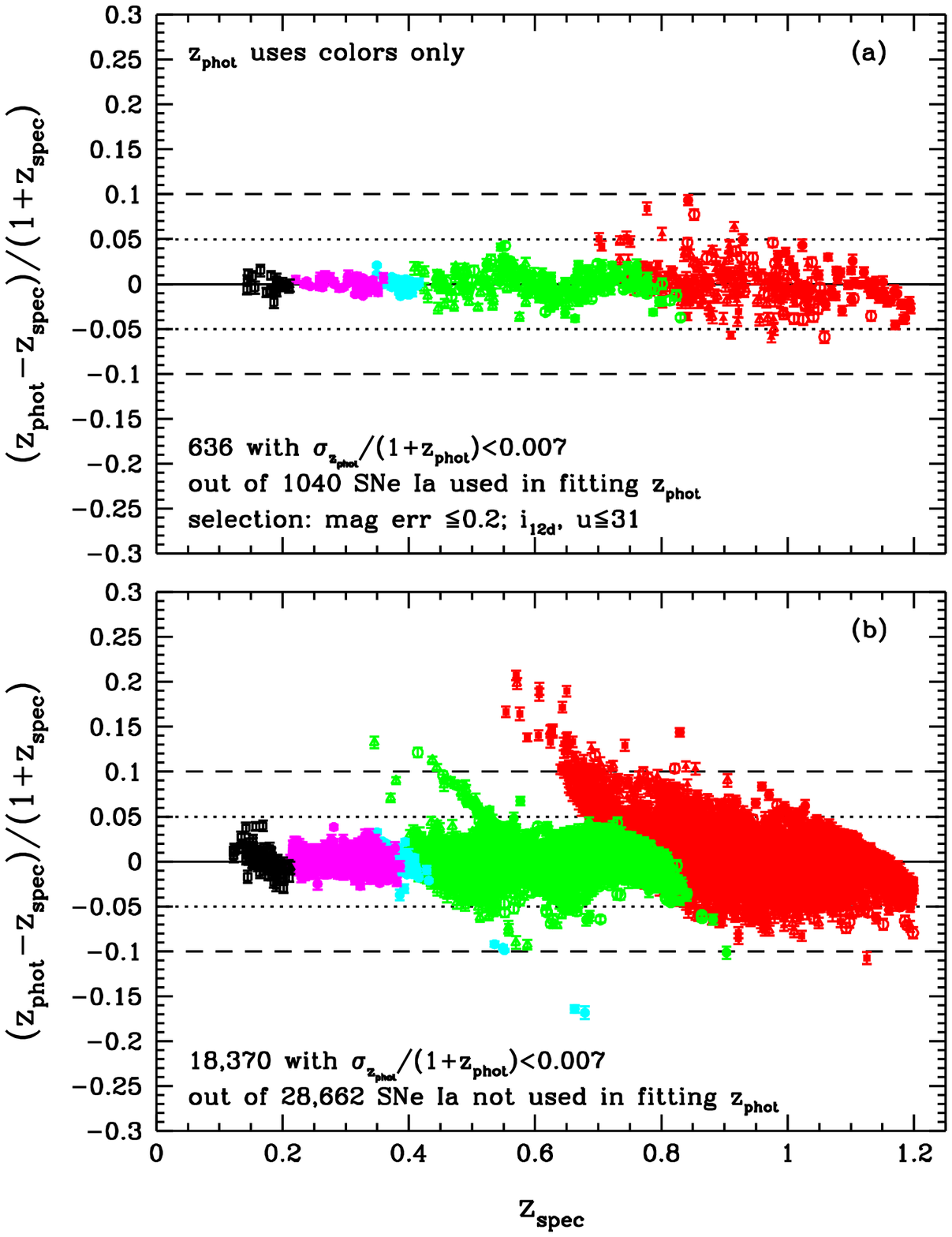}
\caption{The same as Fig.\ref{fig:zaccut0d1}, but with cut 
$\sigma_{z_{\rm phot}}/(1+z_{\rm phot})<0.007$ for improved accuracy, precision, and outlier reduction.}
\label{fig:zaccut0d007}
\end{figure}

Fig.\ref{fig:zaccut0d1}  and Fig.\ref{fig:zaccut0d007} are results corresponding to
Fig.\ref{fig:zacut0d1} and Fig.\ref{fig:zacut0d01}, but for $z_{phot}$ obtained using
Eq.(\ref{eq:zc}), which uses SN Ia colors only.
Using the SN Ia colors only, we obtain a set of photo-z's with similar quality to
using SN Ia color and $i$ band peak flux by
requiring that $\sigma_{z_{\rm phot}}/(1+z_{\rm phot})<0.007$; this leads to a set
of photo-z's with 2 percent accuracy, a bias in $z_{\rm phot}$ of $5.9\times 10^{-4}$,
and an outlier fraction of 0.32\%. The threshold level of 
$\sigma_{z_{\rm phot}}/(1+z_{\rm phot})<0.007$ is set naturally by requiring that
there are no outliers in the training set.
There are 58 outliers (with $\left|(z_{\rm phot}-z_{\rm spec})/(1+z_{\rm spec})\right|>0.1$) 
in the second panel from the top in Fig.\ref{fig:zaccut0d007} (the majority are from the rizY color groups
G11 through G15 --- similar to Fig.\ref{fig:zacut0d01}),  representing 0.32\% of all 18,370 SNe Ia from the test sample that passed this cut.
We do not expect such a small fraction of outliers to have noticeable impact on the cosmological fit; 
we will investigate this in future work.

\begin{figure}
\includegraphics[width=1.2\columnwidth,clip]{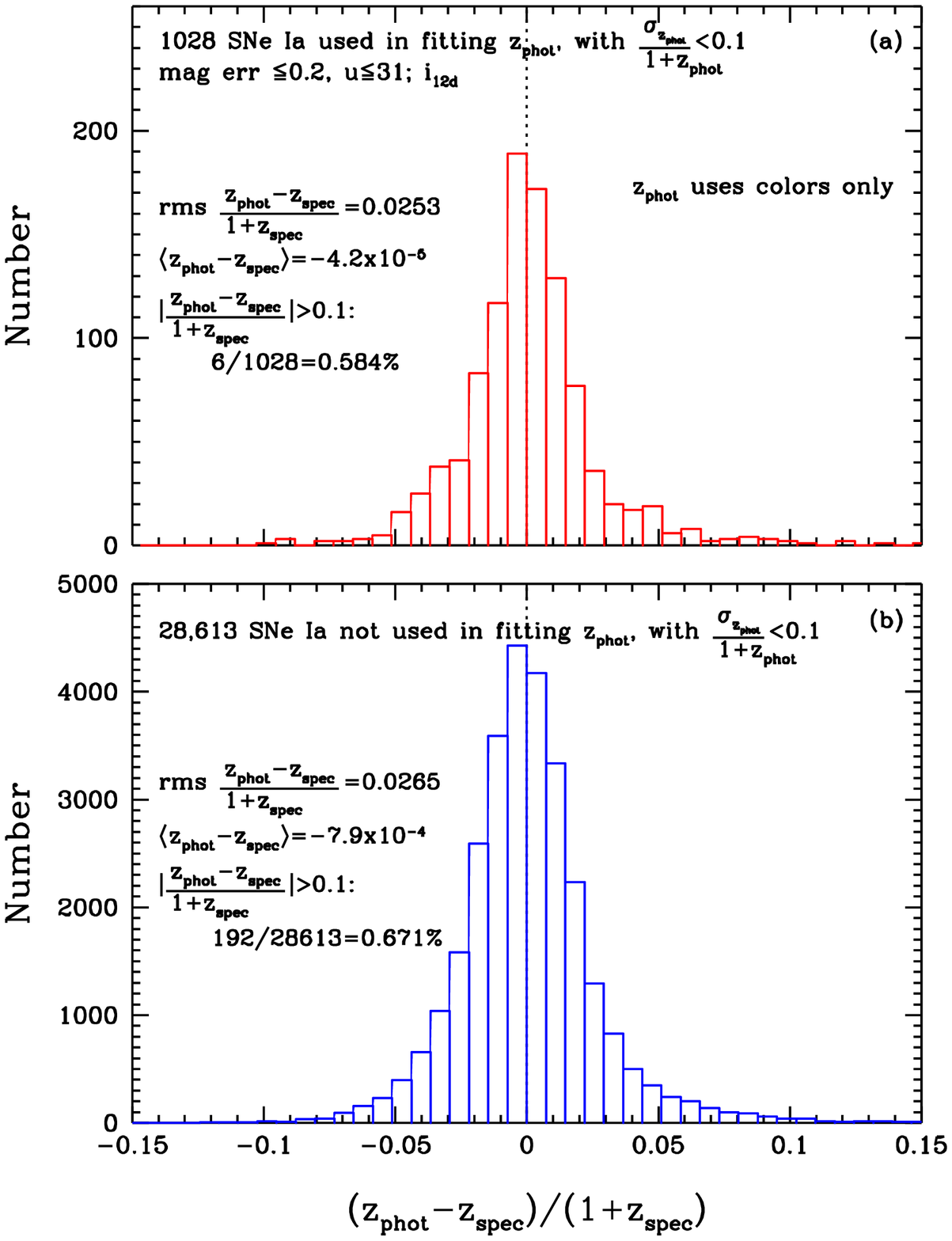}
\caption{The distribution of $(z_{\rm phot}-z_{\rm spec})/(1+z_{\rm spec})$ from Fig.\ref{fig:zaccut0d1}.}
\label{fig:zaccut0d1_bin}
\end{figure}

\begin{figure}
\includegraphics[width=1.2\columnwidth,clip]{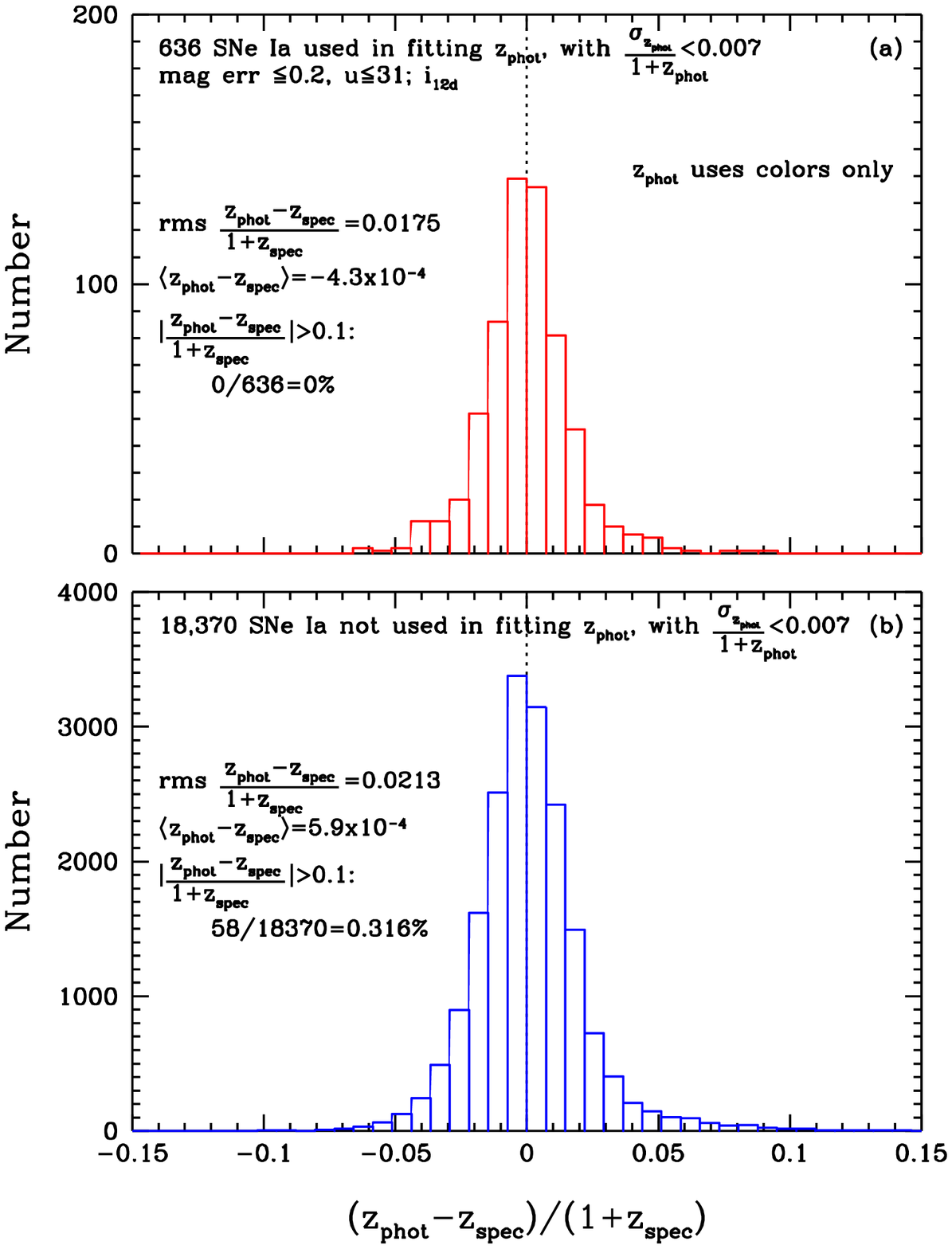}
\caption{The distribution of $(z_{\rm phot}-z_{\rm spec})/(1+z_{\rm spec})$ from Fig.\ref{fig:zaccut0d007}.}
\label{fig:zaccut0d007_bin}
\end{figure}

\section{Summary and Discussion}
\label{sec:sum}

We have shown that a simple analytic photo-z estimator for SNe Ia can be built 
for the LSST, given suitably chosen training sets of SNe Ia with spectroscopic redshifts that
are representative of the general properties of SNe Ia from the LSST.
While our method is based on that proposed by \cite{Wang07}, and advanced by
\cite{Wangetal2007}, it represents a significant improvement over previous work
(which require the use of both colors and $i$ magnitudes of SNe Ia)
in enabling the use of SN Ia colors only. This eliminates the possible complications
that can arise from using SN Ia brightness in estimating photo-z's, since 
the peak brightness of a SN Ia is closely correlated with its distance from us.

Our sample of simulated LSST SNe Ia is representative of data expected from the LSST, based on our current knowledge of SNe Ia.
The multiple-band photometry from the LSST enables us to divide the SNe Ia
into different sets (see Sec.\ref{subsec:LSSTall} and Table \ref{tab:sets}). 
For each set, a photo-z estimator is constructed
using only quadratic functions of colors (see Eq.[\ref{eq:zc}]),
with the constant coefficients given by fitting to 
a training set that contains a maximum of 100 SNe Ia with spectroscopic redshifts
(see Table \ref{tab:sets}).
Note that the division of SNe Ia into color groups is empirical, dependent on the 
properties of the training set, assumed to be representative of the overall sample.

In addition to how the photometric SNe Ia should be grouped, and the performance of our
photo-z estimator for each color group, Table \ref{tab:sets} contains other helpful information as well.
As expected, the performance of the photo-z estimator improves with the increasing number of filters in which 
photometric data are available; note that knowing the filters in which the SNe Ia are visible provides a crude estimate of the
redshift range of the SNe Ia.
Not surprisingly, the performance of the photo-z estimator generally degrades with the increasing redshift of the SNe Ia.

While Table \ref{tab:sets} lists all the SNe Ia that passed our data quality cut (see Sec.\ref{sec:sims}),
Table \ref{tab:cuts} shows us how to derive a high purity sample of photo-z's by making cuts on the
SNe Ia based on the estimated errors of the photo-z's.
This is very useful since we are systematics limited, rather than statistics limited, in using photometric 
SNe Ia for cosmology.

The performance of our simple analytic SN Ia photo-z estimator is quite
impressive when applied to simulated LSST SN Ia photometric data (see Table \ref{tab:cuts}).
We find that the estimated errors on the photo-z's, $\sigma_{z_{\rm phot}}/(1+z_{\rm phot})$, can be used
as filters to produce a set of photo-z's that have high precision, accuracy, and purity.
Using SN Ia colors as well as SN Ia peak magnitude in the $i$ band, we obtain a set of
photo-z's with 2 percent accuracy
(with $\sigma(z_{\rm phot}-z_{\rm spec})/(1+z_{\rm spec}) = 0.02$), a bias in $z_{\rm phot}$
(the mean of $z_{\rm phot}-z_{\rm spec}$) of $-9\times 10^{-5}$, and an outlier
fraction (with $\left|(z_{\rm phot}-z_{\rm spec})/(1+z_{\rm spec})\right|>0.1$) of
0.23 percent, with the requirement that $\sigma_{z_{\rm phot}}/(1+z_{\rm phot})<0.01$.
Using the SN Ia colors only, we obtain a set of photo-z's with similar quality by
requiring that $\sigma_{z_{\rm phot}}/(1+z_{\rm phot})<0.007$; this leads to a set
of photo-z's with 2 percent accuracy, a bias in $z_{\rm phot}$ of $5.9\times 10^{-4}$,
and an outlier fraction of 0.32 percent.

In future work, we will investigate whether the precision and accuracy achieved by our photo-z estimator 
is adequate for cosmology, We will first need to simulatte more realistic samples of LSST SNe, which include
non-Ia's. We will need to expand and advance our photo-z estimator for weeding out non-Ia SNe.
There is much work to do before our photo-z estimator can be applied to the real LSST data.

In principle, our method for estimating photo-z's should be applicable to other surveys, but its performance 
depends on the precision of the photometry of the surveys. However, it can still be used as a quick way
of estimating photo-z's that is model-independent, and complementary to the template-based methods.
The application of our method to non-Ia SNe and galaxies can also be explored. This method for estimating
photo-z's for SNe Ia was initially inspired by a similar method for estimating galaxy redshifts presented
by \cite{Wang98}. It would be interesting to develop a simple photo-z estimator for LSST galaxies based on this
method.

We expect that our SN photo-z estimator for the LSST will be useful in probing
dark energy using LSST SNe Ia with photometry only. It will also provide an
independent cross-check for SN photo-z's derived from other, template-based 
photo-z estimators (see, e.g., \cite{Kim07,Kessler10,Palanque10}).
Our linear photo-z model is very convenient for propagating
photometric uncertainties, especially compared to more ``black-box'' type machine learning algorithms. 
In future work, we will examine the cosmological constraints that can be obtained using LSST SNe Ia with photometry only, with photo-z's estimated using the method presented in this paper.

\section*{Acknowledgments}

We are grateful to Michel Wood-Vasey and Alex Kim for helpful discussions,
and Kirk Gilmore for an internal review of our paper on behalf of
the LSST Dark Energy Science Collaboration.
YW was supported in part by NASA grant 12-EUCLID12-0004.

\setlength{\bibhang}{2.0em}

\clearpage

\begin{table}
\begin{tabular}{|l|l|l|l|l|l|l|l|l|l|l|}
\hline
set & bands & subdivision  &symbol in Fig.\ref{fig:zacut0d1} &$N_s$ & $N_{test}$ & colors & $\sigma\left[\frac{\Delta z}{1+z}\right]$   & $\langle z_{\rm phot}-z\rangle$ & outliers & flagged \\
&&& \& Fig.\ref{fig:zaccut0d1} &&   & only &training, test & training, test & training, test & in test\\
\hline
G11 & $rizY$ & $i\geq 24.8$                & red filled circles &100 & 2904 & N& .0255, $\,$ .0299 & $-$2.6E-7, $\, -$.0074 & 0, 22 & 15\\
    &        &				   &                    &    &      & Y& .0264, $\,$.0261 & $-$1.6E-7, $\, -$.0032 & 0, 8   & 0\\					
G12 & $rizY$ & $i=[24.5, 24.8)$            & red empty circles &100 & 3060 & N&.0308,$\,$ .0280 & $-$8.1E-8, $\, -$.0039 & 1, 11 & 0\\
  &  &             &  & &  & Y&.0313, $\,$ .0280 & $-$6.0E-7, $\, -$.0051 & 1, 14 & 0\\   
G13 & $rizY$ & $i=[24, 24.5)$, $r\geq 25.2$& red filled triangles &100 & 4184 &N& .0274, $\,$.0290 & 2.5E-7, $\, -$.0017 & 0, 25 & 0\\
  &  &             &  & &  & Y&.0280,  $\,$.0289 & $-$1.0E-7, $\, -$8.7E-4 & 1, 26 & 0\\   
G14 & $rizY$ & $i=[24, 24.5)$,  $r<25.2$  & red empty triangles   &100 & 1505 & N&.0301, $\,$ .0369 & 1.1E-7, $\,$ 3.5E-5 & 1, 23 & 4\\
  &  &             &  & &  & Y&.0306, $\,$ .0362 & $-$6.5E-8, $\,$ 6.1E-4 & 1, 22 & 2\\  
G15 & $rizY$ & $i=[23, 24)$ 		   & red filled squares  &100 & 4900 & N&.0279, $\,$ .0346 & 1.1E-7, $\, -$8.5E-4 & 0, 75 & 0\\
 &  &             &  & &  & Y&.0292, $\,$ .0352 & $-$3.2E-7, $\,$ .0021 & 0, 92 & 0\\  
G16 & $rizY$ & $i<23$ 			   & red empty squares   &40 & 53 & N&.0408, $\,$ .0526 & $-$7.1E-8, $\, -$.033 & 1, 4 & 2\\
 &  &             &  & &  & Y&.0494, $\,$  .0612 & $-$2.6E-8, $\, -$.039 & 2, 5 & 3\\  
\hline
G21 & $izY$ & $i\geq 24.5$ & blue filled circles  &20 & 174 & N & .0202, $\,$ .0472 & $-$1.0E-7, $\, -$.0097 & 0, 6 & 0\\
 &  &             &  & &  & Y&.0251, $\,$ .0393 & 1.4E-8, $\,$ .0032 & 0, 3 & 0\\  
G22 & $izY$ & $i< 24.5$    & blue empty circles  &20 & 18 & N&.055, $\,$ .108 & $-$1.3E-8, $\, -$.068 & 1, 7 & 2\\
 &  &             &  & &  & Y&.0627, $\,$ .0798 & 4.3E-8, $\, -$.0508 & 2, 5 & 0\\  
\hline
G31 & $grizY$ & $i\geq 23.5$     & green filled circles  &100 & 2508 &N& .0163, $\,$ .0196 & $-$1.5E-7, $\,$ .0023 & 0, 3 & 2\\
&  &             &  & &  & Y&.0165, $\,$ .0192 & $-$3.0E-8, $\,$ .0035 & 0, 1 & 0\\  
G32 & $grizY$ & $i=[23.1, 23.5)$ & green empty circles  &100 & 3242 & N&.0122, $\,$ .0155 & 2.1E-7, $\, -$.0027 & 0, 2 & 0\\
&  &             &  & &  & Y&.0125, $\,$ .0157 & 1.6E-7, $\, -$.0016 & 0, 3 & 0\\  
G33 & $grizY$ & $i<23.1$         & green empty triangles  &100 & 3888 &N& .0137, $\,$ .0179 & 2.0E-7, $\, -$.0069 & 0, 4 & 0\\
&  &             &  & &  & Y&.0151, $\,$ .0183 & $-$3.8E-8, $\, -$.0025 & 0, 6 & 0\\
\hline
G4 & $griz$ &   & cyan filled circles  & 50 & 480 &N&.0055, $\,$ .0229 & 2.7E-8, $\,$ .0018 & 0, 8 & 6\\
&  &             &  & &  & Y&.0059, $\,$ .0421 & $-$5.7E-8, $\,$ .0088 & 0, 11 & 9\\
\hline
G5 & $ugriz$ &  & magenta filled circles  & 50 & 1361 &N&.0048, $\,$ .0094 & $-$1.9E-8, $\, -$5.2E-4 & 0, 1 & 1\\
&  &             &  & &  & Y&.0063, $\,$ .0123 & 7.2E-8, $\,$ 9.6E-4 & 0, 1 & 0\\
\hline
G61 & $ugri$ & $r\geq 21$ &  black empty triangles  &20 & 75 & N&.0049, $\,$ .048 & $-$1.1E-7, $\,$ .0091 & 0, 4 & 4\\
&  &             &  & &  & Y&.0097, $\,$ .0507 & $-$2.1E-8, $\, -$.010 & 0, 4 & 1\\
G62 & $ugri$ & $r< 21$    &  black empty squares &40 & 310 & N&.0082, $\,$ .026 & $-$2.1E-7, $\,$ .0019 & 0, 1 & 1\\
&  &             &  & &  & Y&.0168, $\,$ .0319 & 8.5E-9, $\,$ .0039 & 0, 6 & 1\\
\hline
\end{tabular}
\caption{Division of SNe into sets according to available photometric passbands, and into
subsets in each set for improved precision. The ``outliers'' column indicates the number of
SNe in the training set and test set that have $\left|(z_{\rm phot}-z_{\rm spec})/(1+z_{\rm spec})\right|>0.1$.
The ``flagged in test'' column indicates the number of outliers in the test set that have
$\sigma_{z_{\rm phot}}/(1+z_{\rm phot})\geq 0.1$. }
\label{tab:sets}
\end{table}

\clearpage

\begin{table}
\begin{center}
\begin{tabular}{|l|l|l|l|l|l|l|l|l|}
\hline
set & $\frac{\sigma_{z_{
m phot}}}{1+z_{
m phot}}<$ & colors only &$N_{tot}$ & $N_{cut}$  & $\sigma\left[\frac{\Delta z}{1+z}\right]$ & $\langle z_{\rm phot}-z\rangle$  & flagged outlier & outliers\\
\hline
training &  100 &N& 1040 & 0 & .0241 & $-$9.6E-8 & 0 & 4 \\
 &   &Y& 1040 & 0 & .0259 & $-$1.8E-6 & 0 & 7 \\
test & 100 &N& 28662 & 0 &  .0270 & $-$.0027 & 0 & 196 (0.68\%) \\
 &  &Y& 28662 & 0 &  .0271 & $-6.9\times 10^{-4}$ & 0 & 207 (0.72\%) \\
\hline
training &  0.1 &N& 1028 & 12 & .0236 & $-$3.1E-5 & 0 & 4 \\
 &   &Y& 1028 & 12 & .0253 & $-$4.2E-5 & 1 & 6 \\
test & 0.1 &N& 28553 & 109  &  .0261 & $-$.0024 & 37 & 159 (0.56\%) \\
 &  &Y& 28613 & 49 &  .0265 & $-7.9\times 10^{-4}$ & 15 & 192 (0.67\%) \\
\hline
training &  0.05 &N& 1011 & 29 & .0229 & $-$.00022 & 1 & 3 \\
 &   &Y& 1017 & 23 & .0242 & $-$5.1E-4 & 3 & 4 \\
test & 0.05 &N& 28327 & 335 &  .0257 & $-$.0021 & 53 & 143 (0.50\%) \\
 &  &Y& 28499 & 163 &  .0262 & $-6.8\times 10^{-4}$ & 31 & 176 (0.62\%) \\
\hline
training &  0.02 &N& 918 & 122 & .0204 & $-$.00073 & 3 & 1 \\
 &   &Y& 949 & 91 & .0221 & $-$3.0E-4 & 5 & 2 \\
test & 0.02 &N& 26606 & 2056 &  .0242 & $-$.0010 & 95 & 101  (0.38\%)\\
 &  &Y& 27480 & 1182 &  .0250 & $-3.0\times 10^{-4}$ & 74 & 133 (0.48\%) \\
\hline
training &  0.01 &N& 705 & 335 & .0173 & $-$.00020 & 4 & 0 \\
 & 0.007  &Y& 636 & 404 & .0175 & $-$4.3E-4 & 7 & 0 \\
test & 0.01 &N& 19640 & 9022 &  .0212 & $-$9.0E-5 & 151 & 45 (0.23\%) \\
 &0.007  &Y& 18370 & 10292 &  .0213 & $5.9\times 10^{-4}$ & 149 & 58 (0.32\%) \\
\hline
\end{tabular}
\end{center}
\caption{Obtaining sets of photo-z's with higher accuracy, precision, and purity
by using the estimated errors on $z_{\rm phot}$ to exclude SNe with large $\sigma_{z_{\rm phot}}/(1+z_{\rm phot})$.}
\label{tab:cuts}
\end{table}



\label{lastpage}


\begin{thebibliography}{}

  \setlength{\itemindent}{-2.5em}

\bibitem[Abell et~al.(2009)]{lsst} Abell, P.~A. et~al. 2009, e-prints arXiv:0912.0201

\bibitem[Albrecht et al.(2006)]{DETF} Albrecht, A., et al., Report of the Dark Energy Task Force; available online at arXiv:astro-ph/0609591v1

\bibitem[Astier et al.(2006)]{Astier06} Astier,  P., et al. 2006, A \& A 447, 31-48 (2006) 

\bibitem[Bernstein et al.(2012)]{bern12} Bernstein, J.P. et al., \emph{Astrophysical Journal} 753 (2012) 152

\bibitem[Caldwell \& Kamionkowski(2009)]{Caldwell09}
Caldwell, R. R., \& Kamionkowski, M., 2009, Ann.Rev.Nucl.Part.Sci., 59, 397

\bibitem[Campbell et al.(2013)]{Campbell13} Campbell, H., et al., 2013, ApJ, 763, 88

\bibitem[Dilday et al.(2008)]{Dilday08} Dilday, B., et al., 2008, ApJ, 682, 262

\bibitem[Frieman, Turner, \& Huterer(2008)]{Frieman08}
Frieman, J., Turner, M., Huterer, D., ARAA, 46, 385 (2008)

\bibitem[Guzzo et al.(2008)]{Guzzo08} Guzzo, L., et al., 2008, Nature, 451, 541

\bibitem[Guy et~al.(2007)]{guy07} Guy, J. et~al. 2007, Astronomy \& Astrophysics, 466, 11

\bibitem[Guy et~al.(2010)]{guy10} Guy, J. et~al. 2010, Astronomy \& Astrophysics, 523, A7

\bibitem[Hlozek et al.(2012)]{Hlozek12} Hlozek, R., et al., 	2012, ApJ, 752, 79
	
\bibitem[Kessler et~al.(2009a)]{kes09} Kessler, R. et~al. 2009a, ApJS, 185, 32

\bibitem[Kessler et~al.(2009b)]{snana} Kessler, R. et~al. 2009b, PASP, 121, 1028

\bibitem[Kessler et~al.(2010a)]{Kessler10} Kessler, R. et~al., 2010, ApJ, 717, 40

\bibitem[Kessler et~al.(2010b)]{Kess10b} Kessler, R.; Conley, A.; Jha, S.; Kuhlmann, S., arXiv:1001.5210 [astro-ph.IM]

\bibitem[Kim \& Miquel(2007)]{Kim07} Kim, A.~G.; Miquel, R., 2007, Astroparticle Physics, 28, 448	

\bibitem[Knox, Song, \& Tyson(2006)]{Knox06} Knox, L.; Song, Y.-S.; Tyson, J. A., 2006, PRD, 74, 023512

\bibitem[Li et al.(2011)]{Li11}
Li, M.; Li, X.-D.; Wang, S.; Wang, Y., 2011, Commun.Theor.Phys., 56, 525

\bibitem[Lupton(1993)]{Lupton93} Lupton, R., 1993, ``Statistics in Theory and Practice'', Princeton University Press

\bibitem[Palanque-Delabrouille(2010)]{Palanque10} Palanque-Delabrouille, N., et al., 2010, A \& A, 514, A63

\bibitem[Perlmutter et al.(1999)]{Perl} Perlmutter, S., et al., \emph{Astrophysical Journal} 517 (1999) 565

\bibitem[Ratra \& Vogeley(2008)]{Ratra08}
Ratra, B., Vogeley, M.~S., 2008, Publ.Astron.Soc.Pac., 120, 235

\bibitem[Riess et al.(1998)]{Riess} Riess, A., et al., \emph{Astron.J.} 116 (1998) 109
 
\bibitem[Sako et al.(2011)]{Sako11} Sako, M., Bassett, B., Connolly, B., et al. 2011, ApJ, 738, 162

\bibitem[Uzan(2010)]{Uzan10} Uzan, J.-P. 2010, General Relativity and Gravitation, 42, 2219
 
\bibitem[Wang, Bahcall, \& Turner(1998)]{Wang98}
Wang, Y.; Bahcall, N.; Turner, E.~L., 1998, AJ, 116, 2081
 
\bibitem[Wang(2007)]{Wang07} Wang, Y., 2007, ApJ Lett., 654, 123  

\bibitem[Wang(2008)]{Wang08} Wang, Y. , 2008, JCAP, 05, 021

\bibitem[Wang(2010)]{Wang10}
Wang, Y., {\it Dark Energy}, Wiley-VCH (2010)

\bibitem[Wang, Narayan, \& Wood-Vasey(2007)]{Wangetal2007} Wang, Y.; Narayan, G.; Wood-Vasey, M., 2007,  MNRAS, 382, 377 (2007) 

\bibitem[Weinberg et al.(2013)]{Weinberg12}
Weinberg, D. H.; Mortonson, M. J.; Eisenstein, D.J.; Hirata, C.; Riess, A. G.; Rozo, E
2013, Physics Reports, 530, 87

\end{thebibliography}
\end{document}